\documentclass[twocolumn,showpacs,amsmath,amssymb,nofootinbib,aps,prd,longbibliography,10pt]{revtex4-1}
\usepackage{dcolumn}
\usepackage{bm}
\usepackage{graphicx}
\usepackage[dvipsnames]{xcolor}

\usepackage{amsthm}

\newcommand{\be}{\begin{equation}}
\newcommand{\ee}{\end{equation}}
\newcommand{\ba}{\begin{eqnarray}}
\newcommand{\ea}{\end{eqnarray}}
\newcommand{\non}{\nonumber}
\newcommand{\n}[1]{\label{#1}}
\newcommand{\eq}[1]{(\ref{#1})}
\newcommand{\hh}{\, ,\hspace{0.25cm}}
\newcommand{\hhh}{\, ,\hspace{0.5cm}}
\newcommand{\BM}[1]{{\mbox{\boldmath $#1$}}}
\newcommand{\ind}[1]{\mbox{\tiny #1}}
\newcommand{\indd}[1]{\mbox{\scriptsize #1}}

\begin{document}

\title{Faraday effect of light caused by plane gravitational wave}
\author{Andrey A. Shoom}
\email{andrey.shoom@aei.mpg.de}
\affiliation{Max Planck Institute for Gravitational Physics (Albert Einstein Institute), Leibniz Universität Hannover, Callinstr. 38, D-30167, Hannover, Germany}


\begin{abstract} 
 
Gravitational field can cause a rotation of polarisation plane of light. This phenomenon is known as the gravitational Faraday effect. We study the gravitational Faraday effect of linearly polarised light propagating in the gravitational field of a weak plane gravitational wave (GW) with ``$+$", ``$\times$", and elliptical polarisation modes. The corresponding gravitational Faraday rotation is proportional to the wave amplitude and to the squared distance traveled by light and it is inversely proportional to the GW's squared wavelength. The rotation is also maximal if light propagates in the direction perpendicular to the GW propagation, along directions of its polarisation. There is no gravitational Faraday rotation when light and the GW propagate in the same or opposite directions, or it propagates along directions perpendicular to directions of the GW polarisation. Helicity of elliptically polarised GW gives higher-order contribution to the gravitational Faraday rotation.

\end{abstract}

\maketitle

\section{Introduction}

The Faraday effect of light is a magneto-optical phenomenon\textemdash rotation of polarisation plane of light propagating in a transparent material in the presence of magnetic field along the light propagation \cite{LL8}. There is analogous effect, the {\em gravitational Faraday effect of light}\textemdash rotation of polarisation plane of an electromagnetic wave propagating in a gravitational field of non-zero angular momentum. Such gravitational field can  be created by a rotating black hole, whose angular momentum causes ``dragging" of inertial frames, which results in rotation of polarisation plane of light and gyroscope precession. These phenomena can be observed in local frames which do not rotate relative to asymptotic rest frame. Study and observation of the gravitational Faraday effect of light can be found in many works \cite{Skrotskii,Plebanski,God,GF2,GF3,GF4,GF5,GF6,CariniRuffini,GF7,GF8,GF9,GF10,Ghosh,Frolov:2011mh,Frolov:2012zn,Shoom:2020zhr,Frolov:2020uhn}. 

It is known that gravitational wave (GW) carries angular momentum. Thus, space-time of a GW can also cause inertial frame dragging, and, as a result, the gravitational Faraday effect. This phenomenon was already studied on the example of linearly polarised electromagnetic shock wave interacting with strong gravitational field of a plane-fronted GW in full nonlinear regime, with different definition of the Faraday rotation angle \cite{Halilsoy:2006ev,Al-Badawi:2010dcp}. Here we consider this effect due to a weak plane gravitational wave in linearised gravity. We study gravitational Faraday rotation caused by a plane GW of ``$+$", ``$\times$", and elliptical polarisation modes.    

This paper is organised as follows. In section II we briefly review the laws of geometric optics in a curved space-time and describe the gravitational Faraday effect of light in the formalism presented in \cite{Shoom:2020zhr} and adopted here to linear polarisation. Section III contains study of the gravitational Faraday effect of light caused by plane gravitational wave. The last section contains  discussion of our results. We shall use a system of units with $c=G=1$ and conventions adopted in the book \cite{MTW}.  

\section{Gravitational \\ Faraday Effect of Light}

When electromagnetic wave (light) propagating in a curved space-time background is highly monochromatic over some space-time regions, we can use an asymptotic short-wave (geometric optics) approximation \cite{MTW}. This approximation is valid when the reduced wavelength (wavelength/$2\pi$) measured in a local Lorentz frame is much less than inhomogeneities in the wave and in the space-time. By using this approximation we consider the electromagnetic wave as a plane-fronted monochromatic wave whose propagation can be described by laws of geometric optics,\footnote{Here and in what follows the semicolon stands for the covariant derivative associated with space-time metric $g_{\alpha\beta}$.}  
\ba
k^{\alpha}k_{\alpha}=0&\hh&k^{\beta}k^{\alpha}_{\,\,\,;\beta}=0\,,\n{go1}\\
p^{\alpha}p_{\alpha}=1&\hh& k^{\alpha}p_{\alpha}=0\hh k^{\beta}p^{\alpha}_{\,\,\,;\beta}=0\,.\n{go2}
\ea
Here $k^{\alpha}=dx^{\alpha}/d\lambda$ is wave vector tangent to a light ray $\Gamma\!\!:x^{\alpha}=x^{\alpha}(\lambda)$, where $\lambda$ is affine parameter of the ray, and $p^{\alpha}$ is the linear polarisation vector.\footnote{As it is in magneto-optical Faraday effect, here we consider light with linear polarisation.} These laws imply that light rays are space-time null geodesics \eq{go1} and the polarisation vector is orthogonal to the light ray and parallel propagated along it \eq{go2}. 

\begin{figure}[htb]
\begin{center}
\includegraphics[width=7cm]{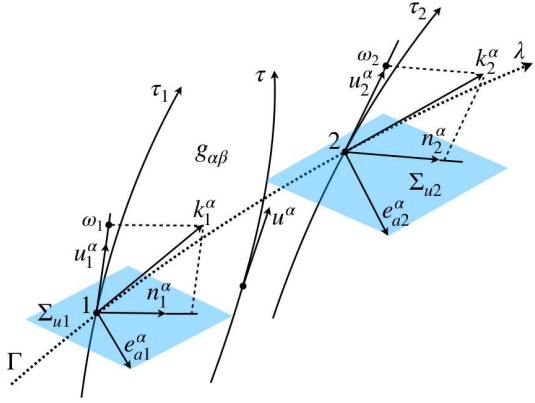}
\caption{World lines of observers in space-time with metric $g_{\alpha \beta}$. The doted line represents world line of light $\Gamma$ with affine parameter $\lambda$. Observers with proper times $\tau_{1}$ and $\tau_{2}$ detect light (events $1$ and $2$) with frequencies $\omega_{1}$ and $\omega_{2}$ propagating in the directions $n_{1}^{\alpha}$ and $n_{2}^{\alpha}$ defined in their local subspaces $\Sigma_{u1}$ and $\Sigma_{u2}$.}\label{fig1}
\end{center}
\end{figure}

Our goal is to study propagation of the polarisation vector along spatial trajectory of a light ray. This can be done by an observer-defined local decomposition of space-time into space and time, which is known as space-time threading approach.\footnote{The threading approach is different from another space-time decomposition known as slicing or the ADM approach. It was originally developed by M\o ller, Zel’manov, and Cattaneo, and discussed in detail in \cite{Jantzen,Carini}. It is considered in \cite{LL2} as well.} Consider a family of observers filling a three-dimensional space like a continuous medium (see Fig.~\ref{fig1}). World lines of these observers form a congruence of integral curves\footnote{Only one such curve passes though each event of the space-time.} of the timelike future directed unit vector field $u^\alpha=u^\alpha(x^\alpha)$, $u^\alpha u_\alpha=-1$. 

\noindent 
Each of the observers defines the local frame of reference, a right-handed orthonormal Lorentz frame:
\ba\n{frame}
\{e_{0}^{\,\,\,\alpha}&=&u^{\alpha},\, e_a^{\,\,\,\alpha}\,;a=1,2,3\}\,,\non\\
u^{\alpha}u_{\alpha}=-1\,,&&e_a^{\,\,\,\alpha}u_{\alpha}=0\hh e_a^{\,\,\,\alpha}e_b^{\,\,\,\beta}g_{\alpha \beta}=\delta_{ab}\,.
\ea
Here $\delta_{ab}$ is the three-dimensional Kronecker tensor. Let $\Sigma_{u}$ be three-dimensional local subspace of the tangent space defined at every event on the observer's world line and orthogonal to $u^{\alpha}$. A vector from the tangent space can be projected into the subspace $\Sigma_{u}$ by the projection operator
\be\n{proj}
p^{\alpha}_{\,\,\,\beta}=\delta^{\alpha}_{\,\,\,\beta}+u^{\alpha}u_{\beta}\,,
\ee
where $\delta^{\alpha}_{\,\,\,\beta}$ is the four-dimensional Kronecker tensor. This operator also defines the induced metric on $\Sigma_{u}$,
\be\n{metric}
p_{\alpha\beta}=p_{\alpha}^{\,\,\,\mu}p_{\beta}^{\,\,\,\nu}g_{\mu\nu}=g_{\alpha\beta}+u_{\alpha}u_{\beta}\,.
\ee
Applying the projection operator to $k^{\alpha}$ we construct the unit spacelike vector $n^{\alpha}$ that defines spatial direction of a light ray. Accordingly, we have
\be\n{k}
k^{\alpha}=\omega(u^{\alpha}+n^{\alpha})\,,
\ee
where $\omega=-k_{\alpha}u^{\alpha}$ is angular frequency of light measured by a local observer (see Fig.~\ref{fig1}).

Let us now consider polarisation vector. As follows from the geometric optics equations, there is the gauge freedom 
\be\n{gauge}
p^{\alpha}\to p^{\alpha}+\kappa k^{\alpha}\,,
\ee
where $\kappa$ is constant along null ray, such that $k^{\alpha}\kappa_{,\alpha}=0$. The gauge transformation implies that at any event on the light world line one can add to the polarisation vector a multiple of the wave vector without affecting the propagation laws \eq{go2}. We fix the gauge so that
\be\n{gauge1}
p^{\alpha}u_{\alpha}=0\,.
\ee
Then, according to the second relation in \eq{go2} and the expression \eq{k} we have
\be\n{trans}
p^{\alpha}n_{\alpha}=0\,.
\ee
Thus, the polarisation vector lies in a two-dimensional subspace perpendicular to $u^{\alpha}$ and $n^{\alpha}$. If we orient the local frame so that $n^{\alpha}=e_3^{\,\,\,\alpha}$, then we can expand the polarisation vector in the local two-dimensional basis $\{e_1^{\,\,\,\alpha}\,, e_2^{\,\,\,\alpha}\}$. To make such an expansion, we note that linear polarisation is a superposition of lefthand and righthand circular polarisation modes whose polarisation vectors rotate in the anti-clockwise and clockwise directions respectively, when viewed along light ray propagation from a particular point on the ray. As in the magneto-optical Faraday effect, due to the gravitational Faraday effect there is an additive contribution $\pm\phi$ to each mode's phase $k_{\alpha}x^{\alpha}$. This contribution results in the net contribution to the linear polarisation mode's phase, which causes rotation of the linear polarisation vector in the anti-clockwise direction by $\phi$.\footnote{For details see Eqs. (95) and (101) in \cite{Frolov:2011mh}.} Taking this into account we have   
\be\n{pol}
p^{\alpha}=\cos\phi\, e_1^{\,\,\,\alpha}-\sin\phi\, e_2^{\,\,\,\alpha}\,.
\ee
\begin{figure}[htb]
\begin{center}
\includegraphics[width=4cm]{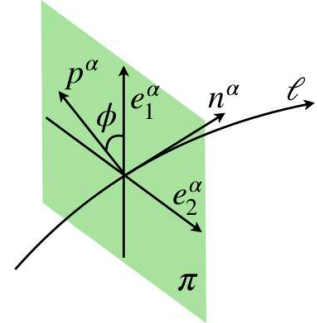}
\caption{Spatial trajectory of a light ray parametrised by its proper length $\ell$. Direction of light propagation is indicated by the unit vector $n^{\alpha}$. The local basis $\{e_1^{\,\,\,\alpha}\,, e_2^{\,\,\,\alpha}\}$ spans two-dimensional plane $\pi$ orthogonal to $n^{\alpha}$. Orientation of the polarisation vector $p^{\alpha}$ lying in the plane is defined by the polarisation phase $\phi$.}\label{fig2}
\end{center}
\end{figure}
\noindent 
We shall call the angle $\phi$ the {\em polarisation phase}. It defines orientation of the polarisation vector (see Fig.~\ref{fig2}). We would like to calculate how the polarisation phase changes along the null ray. Using the third expression in \eq{go2} and the expression \eq{pol} we derive
\be\n{prop1}
k^{\alpha}\phi_{;\alpha}=-e_{1\alpha}k^{\beta}e^{\,\,\,\alpha}_{2\,\,\,;\beta}\,.
\ee
This expression defines change of the polarisation phase along the null ray with respect to the basis $\{e_1^{\,\,\,\alpha},\,e_2^{\,\,\,\alpha}\}$. This change depends on propagation of the basis along the null ray. For example, if the basis is propagated arbitrarily, the polarisation phase change can take arbitrary value, regardless the null ray and gravitational field. Thus, in order to find the polarisation phase change due to the gravitational field only, we need to find out how to propagate the basis along the ray. This problem is analogous to measurement of polarisation in the magneto-optical Faraday effect where we have to align properly optic axes of a polariser and an analyser. For example, if light trajectory is a straight line, we orient the polariser and analyser optic axes parallel to one another and perpendicular to the light trajectory. If light trajectory is a plane curve, we orient their optic axes parallel to one another and perpendicular to the trajectory plane everywhere along the curve. And finally, if the light trajectory is not plane, we propagate the analyser axis from the initial point, where it is aligned with the polariser axis, to the final point along the light trajectory in such a way that it does not rotate with respect to the trajectory during infinitesimally small step of propagation. This procedure generalises the straight and plane trajectory cases. Such a propagation is called Fermi-Walker transport and is defined by vanishing Fermi derivative of the unit vector indicating direction of the analyser optical axis along the unit vector of the light trajectory (see equation \eq{FW} below) \cite{Synge}. Strictly speaking, ``does not rotate with respect to the trajectory" means that the unit vector undergoes only rotation in the plane spanned by the unit vector of the light trajectory and ``acceleration" vector, which defines its rate of change along the curve (see \S 6.5 in \cite{MTW}). For Fermi-Walker transport along a timelike curve the acceleration vector acquires its usual meaning.

Returning to the expression \eq{prop1} we see that it contains derivative of the basis vector along the null ray vector. To define propagation law for the basis along the null ray we shall use the decomposition \eq{k}, which allows to express the basis propagation via its propagations along the vectors $u^\alpha$ and $n^\alpha$,
\be\n{prop2}
k^{\beta}e^{\,\,\,\alpha}_{a\,\,\,;\beta}=\omega(u^{\beta}e^{\,\,\,\alpha}_{a\,\,\,;\beta}+n^{\beta}e^{\,\,\,\alpha}_{a\,\,\,;\beta})\,,
\ee
where $a=1,2$. Thus, we need to define transport law along the observer's world line and along the ray spatial trajectory. We already discussed above how the basis should be transported along the ray trajectory. Namely, its spatial Fermi-Walker derivative along the tangent to the trajectory unit vector $\BM{n}=n^{\alpha}\partial_{\alpha}$ should vanish,
\be\n{FW}
\nabla^{\ind{FW}}_{\indd{\BM{n}}}\BM{e}_{a}\equiv \nabla_{\indd{\BM{n}}}\BM{e}_{a}-\BM{w}(\BM{n},\BM{e}_{a})+\BM{n}(\BM{w},\BM{e}_{a})=0\,.
\ee 
Here $\BM{e}_{a}=e^{\,\,\,\alpha}_{a}\partial_{\alpha}$, $\BM{w}=\nabla_{\indd{\BM{n}}}\BM{n}$ is the mentioned above ``acceleration" vector, the nabla $\nabla$ stands for the covariant derivative associated with the spatial metric $p_{ab}=e_a^{\,\,\,\alpha}e_b^{\,\,\,\beta}p_{\alpha\beta}$, such that $\nabla p_{ab}=0$, and $(.,.)$ stands for a scalar product in the metric $p_{ab}$. This covariant derivative is related to the covariant derivative associated with the space-time metric $g_{\alpha \beta}$ as follows: 
\be\n{dcov}
(\nabla_{\indd{\BM{n}}}\BM{e}_{a})^{\alpha}=p_{\,\,\,\gamma}^{\alpha}n^{\beta}e^{\,\,\,\gamma}_{a\,\,\,;\beta}\,. 
\ee
Note that the orthogonality condition $(\BM{e}_{a},\BM{n})=0$ is preserved by the Fermi-Walker derivative. 

To define propagation of the basis $\{\BM{e}_{a},\,a=1,2\}$ along $u^{\alpha}$ we discuss first the observers congruence $u^\alpha(x^\alpha)$. We shall consider observers whose local frames are ``tied" to Lorentz frame at asymptotically flat space-time region, which is in turn tied to ``fixed stars". Such observers congruence forms a rigid Cartesian latticework where each observer has fixed coordinate position. In stationary space-times such observers are analogous to Killing observers, that is their 4-velocity is proportional to the space-time's timelike Killing vector field. In contrast to such observers, inertial (freely falling) observers are ``dragged" by the local gravitational field together with null rays. These inertial observers cannot detect local gravitational effects. On the other side, the congruence observers detect the ambient gravitational field. The congruence observers move with acceleration $a^{\alpha}=u^{\beta}u^{\alpha}_{\,\,\,;\beta}$ and rotate with angular velocity 
\be\n{vort}
\omega^{\alpha}=\tfrac{1}{2}\varepsilon^{\alpha\beta\gamma\delta}u_{\beta}u_{\gamma;\delta}\,,
\ee
which is also called the vorticity vector of the observers congruence \cite{Hawking-Ellis}. Here $\varepsilon_{\alpha \beta\gamma\delta}$ is the Levi-Civita (pseudo) tensor, which for the observer-adapted frame \eq{frame} is normalised as
\be\n{Lev1}
\varepsilon_{\alpha\beta\gamma\delta}u^{\alpha}e_1^{\,\,\,\beta}e_2^{\,\,\,\gamma}e_3^{\delta}=+1\,.
\ee
The transport law for each observer from the congruence is (see , e.g. \S 13.6 in \cite{MTW})
\be\n{CFW}
u^{\beta}e^{\,\,\,\alpha}_{a\,\,\,;\beta}=(u^{\alpha}a_{\beta}-a^{\alpha}u_{\beta})e_{a}^{\,\,\,\beta}-\varepsilon_{\alpha\beta\gamma\delta}u^{\beta}\omega^{\gamma}e_{a}^{\,\,\,\delta}\,.
\ee 
From \eq{Lev1} with $e_3^{\,\,\,\alpha}=n^{\alpha}$ it follows that
\be\n{Lev2}
\varepsilon_{\alpha\beta\gamma\delta}u^{\alpha}e_1^{\,\,\,\beta}e_2^{\,\,\,\gamma}=n_{\delta}\,.
\ee
Then, using the expressions above we can compute the right-hand side of the expression \eq{prop1} and derive
\be\n{prop3}
k^{\alpha}\phi_{;\alpha}\equiv\frac{d\phi}{d\lambda}=\omega_{\alpha}k^{\alpha}\,.
\ee
This result allows us to find evolution of linear polarisation angle $\phi$ along a null ray $\Gamma\!\!:x^{\alpha}=x^{\alpha}(\lambda)$,
\be\n{Farad}
\phi=\int_{\Gamma}\omega_{\alpha}k^{\alpha}d\lambda\,,
\ee 
which is known as the {\em gravitational Faraday effect of light.}   

\section{Plane gravitational wave}

GW's detections from coalescing compact binaries opened up a new era of observational astronomy and cosmology. Up to date, there are about 90 GW's detections. Sources of the currently detected GW's have luminosity distance in the range from about 40 to about 5300 Mpc. Detected GW's are well modelled by a plane GW solution to the linearised Einstein field equations. In the transverse-traceless (TT) gauge the space-time metric of a monochromatic plane gravitational wave propagating in the $z$-direction has the following form:
\be\n{ttmetric}
ds^2=-dt^2+(1+h_{+})dx^2+2h_{\times}dxdy+(1-h_{+})dy^2+dz^2\,,
\ee
where 
\be\n{modes}
h_{+}=\Re\{A_{+}e^{-i\Omega(t -z)}\}\hhh h_{\times}=\Re\{A_{\times}e^{-i\Omega(t -z)}\}\,.
\ee
Here $A_{+}$ and $A_{\times}$ are constant amplitudes of the ``$+$" and ``$\times$" polarisation modes and $\Omega$ is their constant frequency. 

By the construction, the TT gauge corresponds to a global Lorentz frame $u^{\alpha}=\delta^{\alpha}_{t}$, which is inertial; world line of $u^{\alpha}$ is the space-time \eq{ttmetric} timelike geodesic, as well as any timelike world line with fixed spatial coordinates $x^a=const$. Such geodesics have fixed coordinate separation (see Exercise 35.5 in \cite{MTW}). According to the construction presented in the previous section, to measure gravitational Faraday effect we have to consider non-inertial observers congruence. Therefore, for the given metric \eq{ttmetric} the congruence of inertial observers, $u^{\alpha}(x^{\alpha})=\delta^{\alpha}_{t}\,,\,\, x^a=const$, is not suitable. The reason is that inertial observers cannot measure local gravitational effects, and in particular, detect gravitational Faraday rotation. The best these observers can do is to measure deviation of their world lines, which is constant in the TT gauge. Solution to this problem is to work in the proper frame of one of such observers. As it is known, local geodesic deviation can indeed be detected in the proper reference frame (see \S 35.5 in \cite{MTW}). Here we consider the proper frame of an inertial observer in the       
Fermi normal coordinates (see p. 332 in \cite{MTW} and for details \cite{Manasse:1963zz}),
\ba\n{pf}
ds^2&=&-\left(1+R_{tatb}x^ax^b\right)dt^2-\frac{4}{3}R_{tacb}x^ax^bdtdx^c\\
&+&\left(\delta_{cd}-\frac{1}{3}R_{cadb}x^ax^b\right)dx^cdx^d+{\cal O}(|x^a|^3)dx^{\alpha}dx^{\beta}\,,\non
\ea
where the components of the Riemann tensor $R_{\alpha\beta\gamma\delta}$ are evaluated on the observer's world line $t=\tau$, $x^a=0$, where $\tau$ is the proper time, i.e. $R_{\alpha\beta\gamma\delta}$ depends on $t$ only. 

We now consider congruence of observers $x^a=const$ in the vicinity of the proper frame \eq{pf}. In accordance with \eq{pf}, 4-velocity of the congruence computed up to second order in $x^a$ is
\be\n{u}
u^{\alpha}(x^\alpha)\approx\left(1-\frac{1}{2}R_{tatb}x^ax^b\right)\delta^{\alpha}_{t}\,,
\ee
and its covariant form reads
\be\n{ucov}
u_{\alpha}(x^\alpha)\approx-\left(1+\frac{1}{2}R_{tatb}x^ax^b\right)\delta^{t}_{\alpha}-\frac{2}{3}R_{tacb}x^ax^b\delta^{c}_{\alpha}\,.
\ee
The corresponding vorticity \eq{vort} of the congruence is
\be\n{uvort}
\omega^{\alpha}\approx\varepsilon^{tabc}R_{tbcd}x^d\delta^{\alpha}_{a}\,,
\ee
where we used the Bianchi identity $R_{t[bcd]}=0$. To compute the polarisation phase \eq{Farad} we need to know wave vector $k^{\alpha}$ of a null geodesic of the space-time \eq{pf}. The space-time Christoffel coefficients are of first order in $x^a$. Therefore, the polarisation phase consistent with the metric \eq{pf} should be computed up to first order in $x^a$. This implies that $k^{\alpha}$ should be computed in flat (Minkowski) metric, that is null geodesics passing through the proper frame origin at $\lambda=0$ are
\be\n{null}
x^{\alpha}(\lambda)\approx k^{\alpha}_{0}\lambda\hhh k^{\alpha}_{0}=(\omega_{0},k^{a}_{0})=const\,,
\ee
where $\omega_{0}$ is light frequency and $k^a_{0}$ is three-dimensional wave vector, such that $|k^a|=\omega_{0}$. Then, the polarisation phase can be computed as follows:
\be\n{Faradpf}
\phi=-\int_{0}^{\lambda}e_{abc}R_{tbcd}[t(\lambda)]x^d(\lambda)k^{a}_{0}d\lambda\,,
\ee
where we used the Minkowski metric approximation $\varepsilon^{tabc}\approx-e_{abc}$, where $e_{abc}$ is the three-dimensional Levi-Civita (pseudo) tensor. 

To compute the polarisation phase we need to know the Riemann tensor components. In the linearised theory of gravity, where $g_{\alpha \beta}=\eta_{\alpha \beta}+h_{\alpha \beta}$, with $|h_{\alpha \beta}|\ll1$, we have (see, e.g. p. 438 in \cite{MTW})
\be\n{R}
R_{\alpha \beta\gamma\delta}=\frac{1}{2}(h_{\alpha \delta,\beta\gamma}+h_{\beta\gamma,\alpha \delta}-h_{\alpha \gamma,\beta\delta}-h_{\beta\delta,\alpha \gamma})\,.
\ee
The chosen proper frame \eq{pf} coincides with the TT frame \eq{ttmetric} to first order in $h_{ab}$. Therefore, the Riemann tensor components of the TT frame are the same to first order in $h_{ab}$ as these of the proper frame. Using accordingly the expressions \eq{R} and \eq{modes} we derive
\be\n{Rh}
R_{txzx}=-R_{tyzy}=-\tfrac{1}{2}h_{+,tz}\hh R_{txyz}=R_{tyxz}=\tfrac{1}{2}h_{\times,tz}\,.
\ee
These components are evaluated at $x^a=0$, in accordance with the proper frame construction \eq{pf}. The proper frame quadratic approximation \eq{pf} is valid if $|x^a|$ is sufficiently less than the reduced GW wavelength, $\lambdabar_{\ind{GW}}$.  

In what follows, we first consider ``$+$" and ``$\times$" polarisation modes separately, and then discuss their superposition\textemdash elliptically polarised modes.   

\subsection{``$+$" polarisation}

We can now compute the polarisation phase \eq{Faradpf} due to the plane GW with ``$+$" polarisation. Using the expressions above and defining direction of light ray $k^a_{0}$ in the proper frame by its spherical angular coordinates $\varphi_0$ and $\theta_0$ we derive
\be\n{polpl}
\phi_{+}(t)=-\frac{A_{+}}{2}\sin^2\theta_0\sin(2\varphi_0)[\Omega t\sin(\Omega t)+\cos(\Omega t)-1]\,.
\ee
Here $t$ is the time measured by proper observer. Let time $t$ corresponds to the proper distance $\ell=ct$ ($c$ is the speed of light) of the light propagation from the proper frame origin, then according to the quadratic approximation in \eq{pf}, we should have $\Omega t=\ell/\lambdabar_{\ind{GW}}\ll1$. In this case, the expression \eq{polpl} can be approximated as follows:
\be\n{polpla}
\left.\phi_{+}(\ell)\right\vert_{\ell\ll\lambdabar_{\ind{GW}}}\approx-\frac{A_{+}}{4}\frac{\ell^2}{\lambdabar_{\ind{GW}}^2}\sin^2\theta_0\sin(2\varphi_0)\,.
\ee

\subsection{``$\times$" polarisation}

The polarisation phase evolution due to the plane GW with ``$\times$" polarisation is
\be\n{polx}
\phi_{\times}(t)=\frac{A_{\times}}{2}\sin^2\theta_0\cos(2\varphi_0)[\Omega t\sin(\Omega t)+\cos(\Omega t)-1]\,.
\ee
According to the quadratic approximation in \eq{pf}, for $t$ corresponding to the distance $\ell$ of the light propagation from the proper frame origin, we should have $\ell\ll\lambdabar_{\ind{GW}}$. In this case, the expression \eq{polx} can be approximated as follows:
\be\n{polxa}
\left.\phi_{\times}(\ell)\right\vert_{\ell\ll\lambdabar_{\ind{GW}}}\approx\frac{A_{\times}}{4}\frac{\ell^2}{\lambdabar_{\ind{GW}}^2}\sin^2\theta_0\cos(2\varphi_0)\,.
\ee

\subsection{Elliptical polarisation}

Elliptical polarisation mode can be presented as a superposition of the ``$+$" and ``$\times$" polarisation modes of different amplitudes,
\be\n{ellmode}
h_{\sigma}=\Re\{[A_{+}+i\sigma A_{\times}]e^{-i\Omega(t -z)}\}\,.
\ee
Here $\sigma=\pm 1$ and $+$($-$) sign stands for the right-(left-) handed polarisation when viewed from the source along the wave propagation. The polarisation phase evolution due to the elliptically polarised plane GW is
\ba\n{ellpol}
\phi_{\sigma}(t)&=&-\frac{A_{+}}{2}\sin^2\theta_0\sin(2\varphi_0)[\Omega t\sin(\Omega t)+\cos(\Omega t)-1]\non\\
&+&\frac{\sigma A_{\times}}{2}\sin^2\theta_0\cos(2\varphi_0)[\sin(\Omega t)-\Omega t\cos(\Omega t)]\,.\non\\
\ea
As in the cases above, according to the quadratic approximation in \eq{pf}, we should have $\ell\ll\lambdabar_{\ind{GW}}$. Then the expression \eq{ellpol} can be approximated as follows:
\ba\n{ellpola}
\left.\phi_{\sigma}(\ell)\right\vert_{\ell\ll\lambdabar_{\ind{GW}}}&\approx&-\frac{A_{+}}{4}\frac{\ell^2}{\lambdabar_{\ind{GW}}^2}\sin^2\theta_0\sin(2\varphi_0)\non\\
&+&\frac{\sigma A_{\times}}{6}\frac{\ell^3}{\lambdabar_{\ind{GW}}^3}\sin^2\theta_0\cos(2\varphi_0)\,.
\ea
We see that the helicity $\sigma$ contribution to the gravitational Faraday rotation is of higher order. 
 
\section{Discussion}
 
Here we studied the gravitational Faraday effect of light due to a weak plane gravitational wave with ``$+$", ``$\times$", and elliptical polarisation modes. We computed the polarisation phase in the proper frame, which is valid for $\ell\ll\lambdabar_{\ind{GW}}$, where $\ell$ is distance from the proper frame origin. The derived expressions show that the gravitational Faraday rotation is proportional to the GW amplitude. It is maximal for electromagnetic wave (light) propagating in the direction perpendicular to the GW propagation $\theta_{0}=\pi/2$, along directions of its polarisation: $\phi_{0}=0,\pi/2$ for ``$\times$" polarisation mode, and $\phi_{0}=\pi/4, 3\pi/4$ for ``$+$" and elliptical polarisation modes. Within the order of approximation, there is no gravitational Faraday rotation when the light and GW propagate in the same directions, or/and the light propagates along directions perpendicular to directions of the GW polarisation. Considering GW of elliptical polarisation we also found that the GW helicity $\sigma=\pm 1$ contribution to the gravitational Faraday rotation is of higher order in $\ell/\lambdabar_{\ind{GW}}$. It would be interesting to analyse the actual amount of the expected polarisation rotation for observed GW, which we may consider elsewhere.       

\bibliography{biblio.bib}

\end{document}